# Fooling with facts:
# Quantifying anchoring bias through a large-scale online experiment

Taha Yasseri[1,2]* and Jannie Reher[1]

[1]Oxford Internet Institute, University of Oxford, 1 St Giles, Oxford OX1 3JS, UK
[2]Alan Turing Institute, 96 Euston Rd, London NW1 2DB, UK

*Corresponding Author: taha.yasseri@oii.ox.ac.uk

## Abstract

Living in the 'Information Age' means that not only access to information has become easier but also that the distribution of information is more dynamic than ever. Through a large-scale online field experiment, we provide new empirical evidence for the presence of the anchoring bias in people's judgment due to irrational reliance on a piece of information that they are initially given. The comparison of the anchoring stimuli and respective responses across different tasks reveals a positive, yet complex relationship between the anchors and the bias in participants' predictions of the outcomes of events in the future. Participants in the treatment group were equally susceptible to the anchors regardless of their level of engagement, previous performance, or gender. Given the strong and ubiquitous influence of anchors quantified here, we should take great care to closely monitor and regulate the distribution of information online to facilitate less biased decision making.

Heuristics are mental shortcuts that enable us to arrive at solutions to complex tasks or problems with minimal effort (Tversky & Kahneman, 1983; Shah & Oppenheimer, 2008). However, as has been shown by Tversky and Kahneman (1974) these shortcuts come at a cost: in order to be able to quickly solve a problem, certain information will be simplified, some ignored, and estimations will be made, thus increasing the likelihood of systematic errors in decisions. Commonly referred to as cognitive biases, these errors are the result of non-rational information processing (Lieder, Griffiths, Huys & Goodman, 2018; Haselton, Nettle & Murray, 2016). Examples of cognitive bias include the anchoring effect (that is, the influence on decisions of the first piece of information encountered), the availability heuristic (where estimates of the probability of an outcome depend on ease of access to that information), the framing effect (the presentation of identical information in different ways), and confirmation bias (a focus on information that supports a pre-existing position), among others. Given the black box nature of the algorithms that drive searching and selecting relevant information on the Internet (Pasquale, 2015), it is left to large Internet corporations managing those algorithms to decide which information is "valuable" and therefore displayed to users (Schroeder, 2014). Recent experiments by Nikolov, Lalmas, Flammini, and Menczer (2018) and Kramer, Guillory, and Hancock (2014) show that depending on what online platform is used, people may be exposed to extremely different information, which can in turn result in the development of "social bubbles" or even changes in their emotional state.

This paper examines one of these effects: anchoring. Tversky and Kahneman (1974) observed that in situations where people make estimates or predictions (e.g. the value of a car), the resulting judgment tends to be similar to a previously encountered value (e.g. what the salesperson is offering). The term 'anchoring' can therefore be understood as people's tendency to rely heavily on these prior values (or 'anchors') when making decisions. This effect has been found to be pervasive and robust in a variety of experimental settings (see Furnham & Boo, 2011; Chapman & Johnson, 2002; Lieder et al., 2018; Bahník, Englich, & Strack, 2017) and real-world contexts, including in courtroom sentencing (Englich, Mussweiler, & Strack, 2006), in negotiations (Galinsky & Mussweiler, 2001; Thorsteinson, 2011), in financial market decisions (Collins, De Bondt & Wärneryd, 2018), in property pricing (Northcraft & Neale, 1987), and in judging the probability of the outbreak of a nuclear war (Plous, 1989). Typically, anchoring bias occurs when numeric anchors are provided, although some research has also investigated the effect of non-numeric anchors (LeBoeuf and Shafir, 2006; Wesslen et al., 2018).

Most experiments investigating anchoring effects have used artificial, laboratory-like conditions. Subjects have been asked to estimate the percentage of African countries in the United Nations (Tversky & Kahneman, 1974), the length of the Mississippi river (Jacowitz & Kahneman, 1995), Gandhi's age at death (Strack & Mussweiler, 1997), and the number of calories in a strawberry (Mochon & Frederick, 2013). Few of these questions, however, seem particularly engaging or representative of everyday situations. Moreover, participants were often incentivised to take part by means of monetary reimbursement or course credit. Such a setting does not encourage participants to concentrate fully on the task at hand, given pay is frequently fixed-rate and non-performance-related. The fact that candidates were often recruited from university courses or online portals like Amazon Mechanical Turk (Mochon & Frederick, 2013; Smith, Windschitl & Bruchmann, 2013) adds further to the artificiality of the experiments. Many experiments only recruited between 30 and 50 participants, which limits the statistical power of the experiments and generalizability of the results. Lastly, most of these experiments did not systematically quantify and generalise the anchoring effect by diversifying and analysing a large range of questions at the same time, or adequately control for subjects already knowing the correct answer to a general-knowledge question.

We address these shortcomings through a large-scale online field experiment undertaken on a mobile game ("Play the Future") which "gives players the power to predict the everyday outcomes of the world's most popular brands and events for points, prestige and prizes". Players make numeric predictions regarding the outcome of certain natural, economic, social, sports and entertainment events such as the weather, stock prices, and flight times (*Figure 1*), in order to receive points and accuracy feedback for each prediction, which they can then compare with other users on the app. We know that predictions about future events rely on a complex process in which prior knowledge is used to determine the likelihood of a given event. However, due to time or knowledge constraints, people rely on heuristics or mental shortcuts when making predictions – so-called "bounded rationality" (Simon, 1972). By systematically manipulating numeric hints provided as part of the game (i.e. manipulating information that is potentially relevant for the judgment process), we examine the influence of anchoring bias on the decisions made by our experimental subjects.

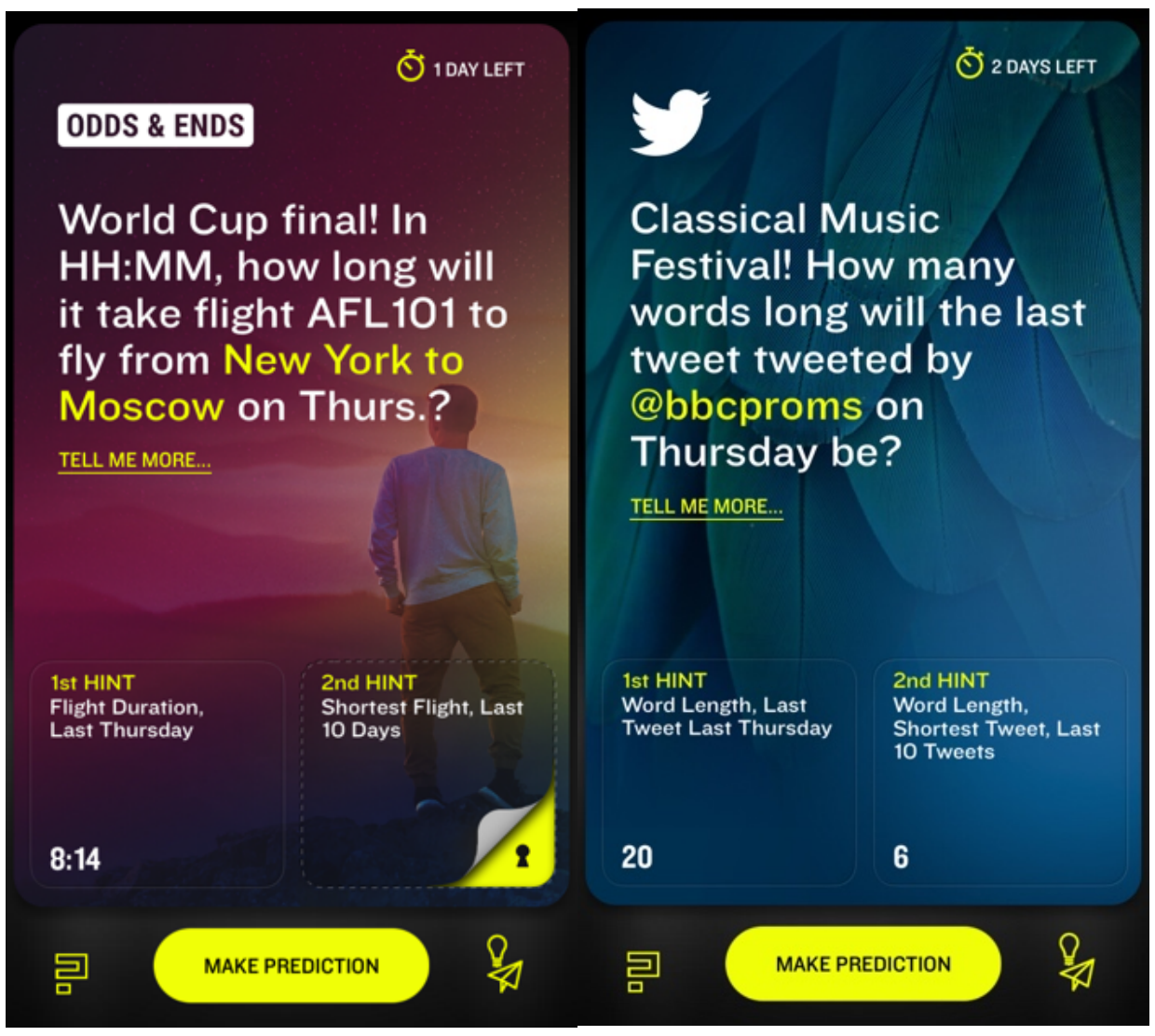

**Figure 1.** Screenshots of two example questions during the experiment on the Play the Future mobile application. The first hint is provided by default. The second hint has been unlocked in the right screenshot.

For the experiment we designed 62 questions (and corresponding hints) to be provided the gamers (see *Table 1* for example questions). The subjects who answered our questions were not explicitly recruited to the experiment, and did not know they were part of the experiment. For each question, two hints are provided that contain information regarding the topic of the question. The first hint is always shown, and is identical for all users. The second hint can be unlocked with a 'key'. Keys automatically renew themselves every four hours for all users for free. We tried to ensure our questions (and the hints) were typical of the game, to ensure they played on the app normally, in a real-world setting. The hints provided were accurate values, i.e. while we varied the type of information available to participants, all information provided was correct.

**Table 1.** Examples of experiment questions.

| | **Control Group** | **Treatment Groups** | | |
|---|---|---|---|---|
| **Question** | **Hint 1 (identical for all users)** | **Hint 2 Group A** | **Hint 2 Group B** | **Result** |
| In HH:MM, how long will it take flight AC880 to fly from Toronto Pearson to Paris CDG on Thursday? | Flight duration, last Thursday: 06:36 h | Longest flight duration during 10-day period: 06:51 h | Shortest flight duration during 10-day period: 06:23 h | 07:04 h |
| In USD, what will Tesla's (TSLA) stock price be at the Nasdaq Stock Market next Tuesday (July 3) at 1 PM? | TSLA's stock price last Tuesday at 1 PM: 334.34 USD | TSLA's stock price low last Tuesday: 326.00 USD | TSLA's stock price high last Tuesday: 343.55 USD | 310.86 USD |

The experimental treatment was administered in the second (locked) hint. Participants who only saw the first hint are considered to be in the control group (for more details see Methods). Users who opted to unlock the second hint were randomly assigned to Group A or Group B, who were then provided with the lowest or highest value of a previous outcome (*Table 1*). We then examined how the provision of both low and high anchors (in the form of factual hints) affects the values predicted by the users. If anchoring is effective, we would expect those provided with a high figure, to predict higher values than those who were provided with low figures. In addition to investigating a general anchoring effect of information on decision making, in an additional set of experiments we provided non-factual hints in the form of "<someone> from Play the Future Team's prediction" to compare the size of the anchoring effect induced by "expert's opinion" to the one by pure factual hints. Finally, we build on research on individual differences in susceptibility to anchoring bias (Eroglu & Croxton, 2010) to examine how the bias varies for individual users according to their level of engagement while playing the game, their previous performance, and their gender.

# Results

## General Anchoring Effect

We asked 42 questions in six different topics over a period of a month. Each question was answered by an average of 219 users, with an average of 58 participants in the treatment condition (i.e. Groups A and B) for each question. The full list of questions and hints is provided in *Supplementary Table 1*. The actual results (i.e. the correct answers) have no effect on the experiment or analysis, given they were known and revealed only after the experiment was over. The answer distributions to the questions shown in *Table 1*, are shown in *Figure 2*. Yuen's test for independent sample means with 15% trimming was applied to all questions to test whether the predictions in the two treatment groups are different from each other. The results indicate that the answer distributions are indeed significantly different ($p < 0.01$) in all questions except one, which is marginally significant ($p = 0.05$).

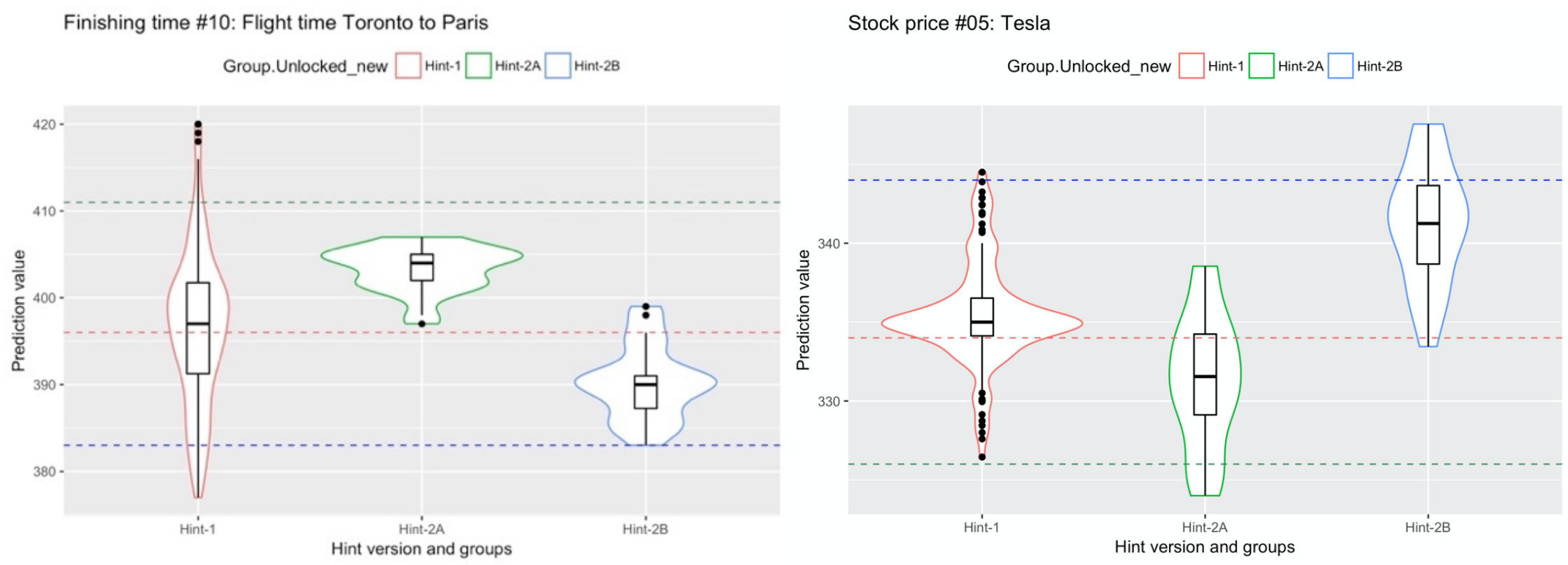

**Figure 2.** Distribution of participants' answers depending on group allocation for the two example questions in *Table 1*. Provided anchors in each group indicated by dotted lines. Diversity of predictions in the two treatment groups is smaller in the left graph compared to the right.

Eight questions providing irrelevant information in the first hint were included in the experiment as a check to determine whether control group participants are blindly following the values provided in the first hint without questioning their actual relevance for the target outcome (see the questions in *Supplementary Table 1)*. If this were the case, it would be inferred that the cognitive effort invested by participants in this game is minimal and that the observed responses are not actually due to anchoring effects but to users' laziness. The detailed results of this test are provided in *Supplementary Figures 1-2*, but in short, it was confirmed that the players of the game, generally pay attention to the relevance of the hint to the question and do not follow the values in the hint blindly.

For 42 standard questions the anchoring stimulus (e*quation (1))* and the anchoring response (*equation (2))* were calculated and shown in *Figure 3*.

(1) $$Stimulus = \frac{|\,Hint\ Group\ A - Hint\ Group\ B\,|}{\sigma_{Control\ Group}}$$

(2) $$Response = \frac{|\,Median\ answer\ Group\ A - Median\ answer\ Group\ B\,|}{\sigma_{Control\ Group}},$$

where, $\sigma$ is the standard deviation of all the answers in each group.

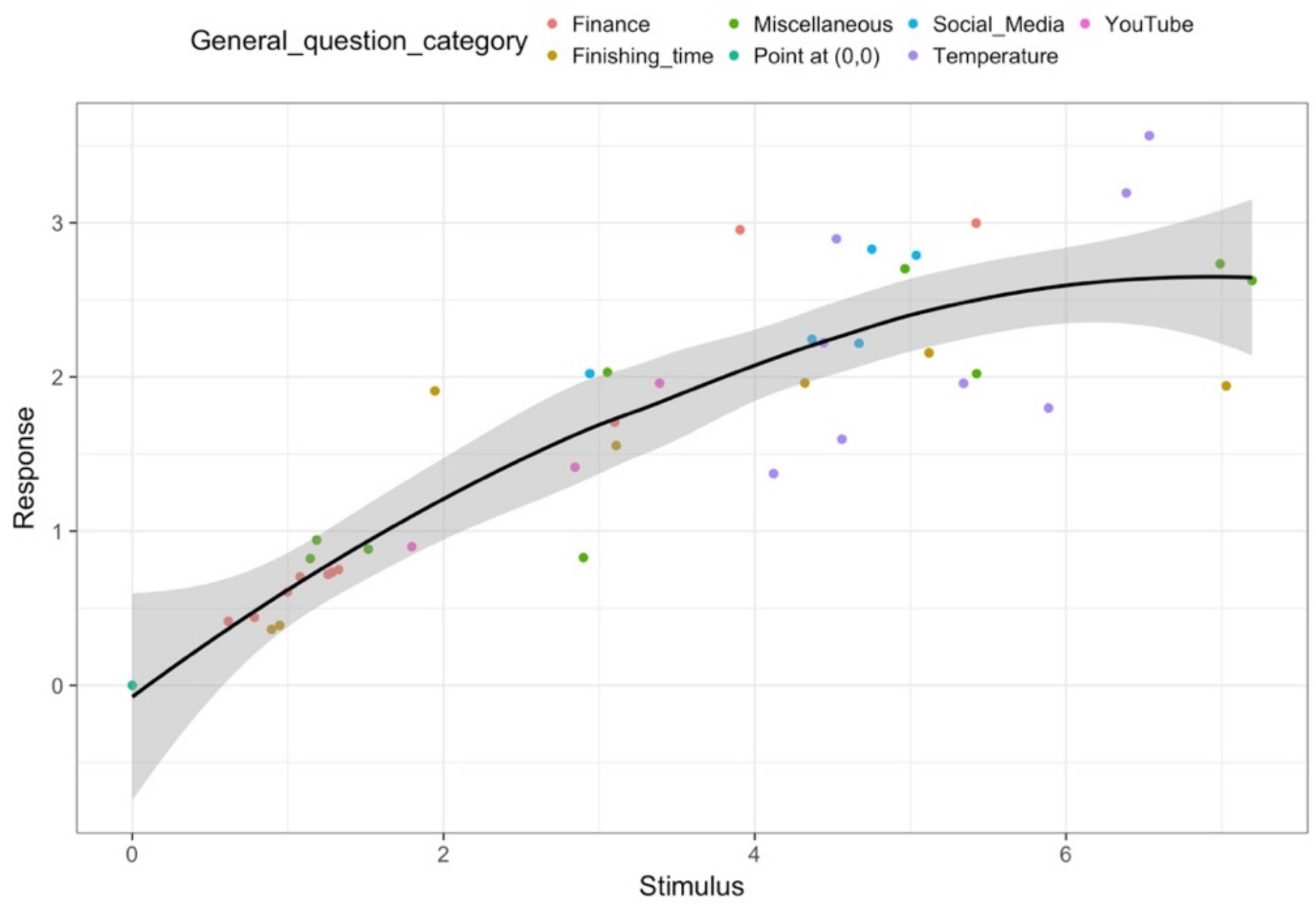

**Figure 3.** Anchoring response function based on 42 standard questions. LOESS curve fitted to all data points. Points coloured according to question category.

As expected, larger stimulus leads to larger anchoring response, however, after an initial increase in the size of the induced anchoring effect, saturation appears.

A closer look at examples shown in *Figure 2* however reveals that in some cases the diversity of answers in the treatment groups is very small (left panel) whereas in other cases the answers are widely dispersed (right panel). This observation suggests that for some questions the provision of two anchors lead to higher collective prediction certainty when compared to the control group, while for other questions it introduced more doubt about the true value of the likely outcome among the group members.

This observation warrants a more systematic analysis of the ratio of the treatment to the control group's standard deviation depending on the size of the provided stimulus. *Figure 4* illustrates how the relative group diversity – that is the diversity of answers within each treatment group divided by the diversity of predictions in the control group (*equation (3)*) – changes with the group stimulus (*equation (4)*).

(3) $$Relative\ group\ diversity\ X = \frac{\sigma_{Group\ X}}{\sigma_{Control\ Group}}$$ , where X = A, B

(4) $$Group\ stimulus = \frac{|\,Hint\ Group\ X - Median\ answer\ Control\ Group\,|}{\sigma_{Control\ Group}}$$ , where X = A, B

Smaller anchoring stimuli lead to smaller relative group diversity, which means that treatment group users were collectively more certain of their answers compared to the control group for these questions. When the size of the anchoring stimulus increases, the relative group diversity also increases.

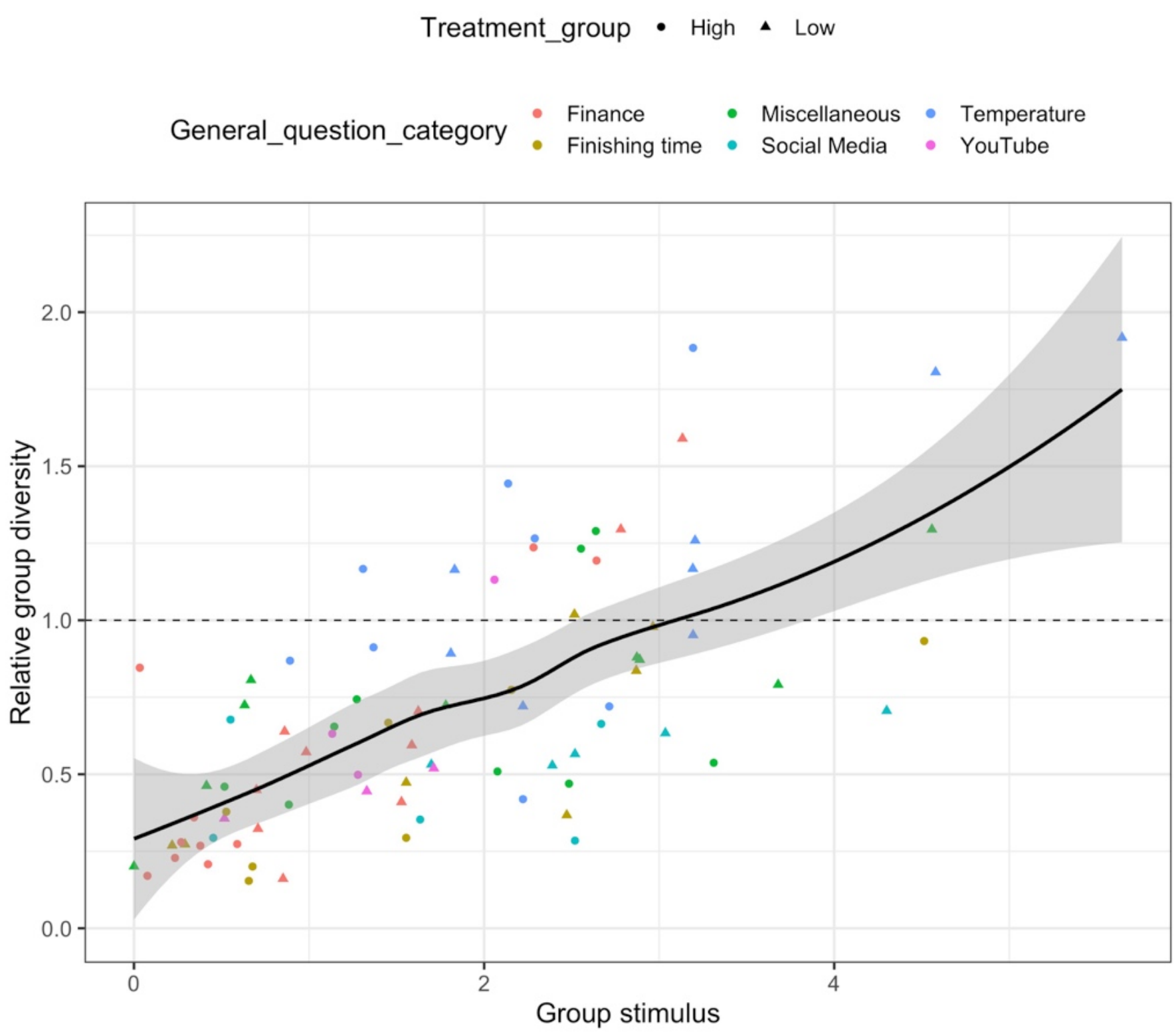

**Figure 4.** Relative group diversity vs. group stimulus. Points coloured according to question category and shaped according to group allocation (high vs. low anchor). Line at y = 1 indicates the equal ratio of the treatment to the control group's standard deviation.

Based on this observation, we define a new measure of the anchoring bias by normalising the difference between the median answers of the two treatment groups by the average of the standard deviations of the two treatment groups instead of the standard deviation of the control group (*equation (5)*).

(5) $$Modified\ response = \frac{|\,Median\ answer\ Group\ A - Median\ answer\ Group\ B\,|}{\left(\frac{\sigma_{Group\ A} + \sigma_{Group\ B}}{2}\right)}$$

The resulting modified response accounts for the participants' collective confidence with regard to their predictions. A small average standard deviation of the two treatment groups is assumed to be indicative of higher certainty of answers among participants since users appear to be in collective agreement regarding the true value of the target. The modified response function is shown in *Figure 5*.

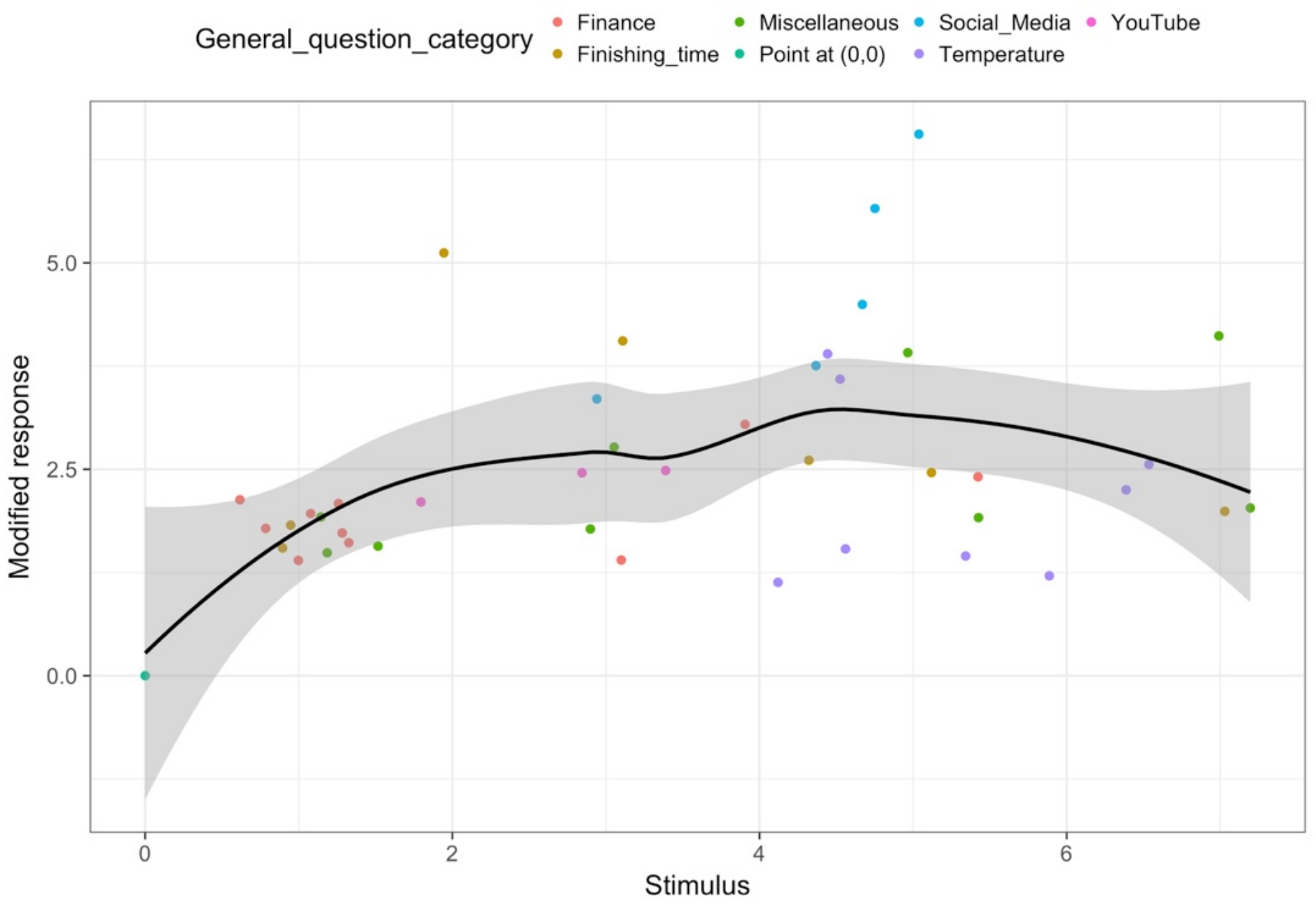

**Figure 5.** Modified response vs. stimulus. Points coloured according to question category.

Medium sized stimuli (2 < x < 5) seem to have caused majority of participants to believe that the anchors might be plausible, resulting in a larger modified response, i.e., larger difference in the answers of the two groups with higher collective confidence in each group. High anchors induce more uncertainty among participants: not all users follow high anchoring stimuli, instead, a growing proportion of participants starts adjusting their predictions to less extreme values, thus increasing the diversity of answers.

## “expert-opinion” anchors

By means of substituting the factual information in the second hints by fictitious “Play The Future-prediction” values (which we carefully selected to resemble realistic predictions) in additional 12 questions, we examined how strongly these values impact participants’ predictions. The hints of this questions presented as the prediction of a hypothetical member of the Play The Future team. An example is given in *Table 2* and the rest of the questions are in *Supplementary Table 1.*

**Table 2.** Example of experiment question with fictitious hints.

| | **Control Group** | **Treatment Groups** | | |
|---|---|---|---|---|
| **Question** | **Hint 1 (identical for all users)** | **Hint 2 Group A** | **Hint 2 Group B** | **Result** |
| In USD, what will Sotheby's (BID) closing stock price be at the NYSE on Tuesday? | Closing stock price last Tues.: 59.06 USD | Sally from PTF's prediction: 60.31 USD | Sally from PTF's prediction: 57.81 USD | 55.84 USD |

Upper panel *of Figure 6* shows the size of the anchoring effect versus the anchoring stimulus for 42 standard and 12 PTF-prediction style questions. It becomes immediately apparent that the questions containing PTF-prediction values in the treatment hints result in consistently larger responses compared to the standard questions with factual information. Middle panel of *Figure 6* shows that not only the medians of treatment groups distributions are moved further with the fictitious hints, but also the relative group diversity tends to be lower for PTF-prediction questions, which is the result of less variation in the treatment groups’ answers compared to the control group's predictions. Hence, in the modified response curve is shown in lower panel of *Figure 6* we see an even larger amplification of the anchoring effect resulted from the fictions hints.

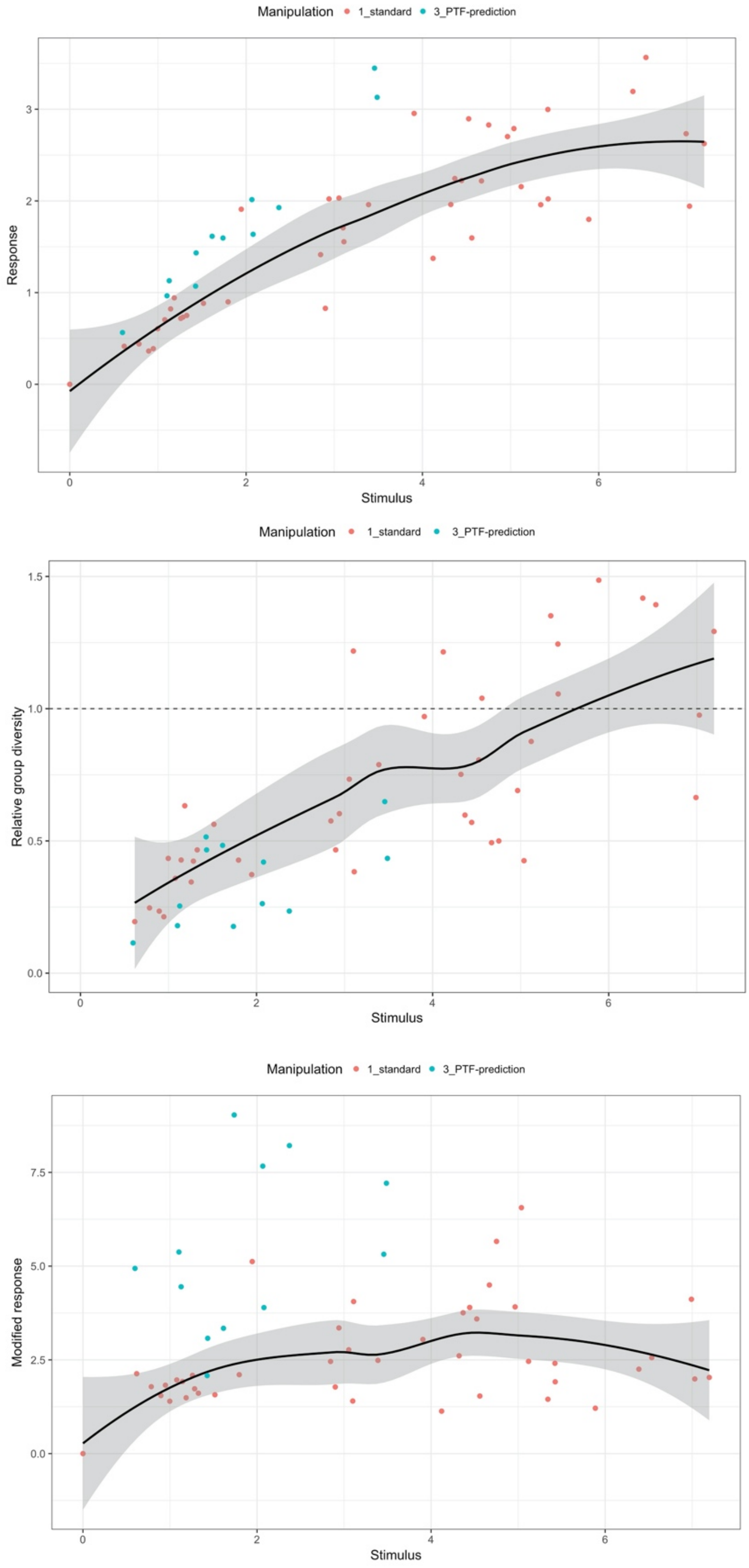

**Figure 6.** *Upper panel*: Response vs. stimulus. *Middle panel:* Relative group diversity vs. stimulus. *Lower panel:* Modified response vs. stimulus for 42 standard and 12 PTF-prediction questions. LOESS curve only fitted to standard anchoring questions for reference.

## Individual analysis

In order to analyse the effect of the anchoring stimulus provided in the experiment on each participant's predictions, the median bias for each user during the experiment was calculated. Firstly, the normalized difference between the user's prediction and the control group's median prediction was computed for each question (*equation (6)*).

(6) $$Bias\ per\ participant\ i\ per\ question = \frac{|\ Prediction_i - Median\ answer\ Control\ Group\ |}{\sigma_{Control\ Group}}$$

Next, the median bias was calculated per individual user for all questions answered in the control condition and for all questions answered in the treatment condition. A higher individual bias indicates that a certain user's prediction values are further away from the median of the control group's predictions, which may be the result of the influence of the anchoring stimuli.

**User engagement:** It could be hypothesized that the anchoring effect is stronger in less experienced users. We therefore tested for a difference between participants who answered less than half of all experimental questions (casual users) and those who made predictions for more than half of the questions ("loyal users"). Results are shown in *Supplementary Figure 3.* Focusing on control groups only, we observe that loyal users in the control group seem to make predictions that resemble the median control group's predictions more closely than casual users in the control group (Welch's $t$-test significant $t = 6.986$, $p < 0.001$). The highly engaged users may have concluded that making moderate rather than extreme predictions (if no further information in the form of a second hint is provided) constitutes a relatively successful strategy in this game.

However, among the treatment groups, barely any difference between casual and loyal participants can be detected (Welch's unequal variances $t$-test insignificant $t = 1.448$, $p = 0.149$). This implies that all users regardless of their level of engagement on the app are roughly equally susceptible to the provided anchoring stimuli. Thus, it is concluded that even among high frequency players no 'learning effect' regarding the true purpose of this experiment occurred.

**User prior accuracy:** Many of the experiment participants had already played the game and the records of their predictions were available to us. We divided the players into high accuracy and low accuracy groups based on their accuracy score in all the games they had played before our experiment and compared the induced bias for the two groups (see Methods for details).

Considering only control groups, participants who made less accurate predictions before the start of the experiment provided answers that were relatively close to the overall median answer during the experiment (*Supplementary Figure 4*). Previously better performing users made more distinct predictions, potentially because they put less trust in the information provided in the first hint. The results of Welch's $t$-test indicate that the individual biases in the two control groups are slightly different from each other ($t(139.11) = 1.933$, $p = 0.055$), however, this result is statistically significant only at the 0.1-level.

For previously low-performing users in the treatment group, the individual bias seems to be slightly larger compared to the bias observed for previously well-preforming users. This would imply that their answers were slightly more influenced by the anchoring stimuli compared to better performing users. However, Welch's $t$-test reveals that the visually observed difference is not statistically significant ($t(97.99) = -1.443$, $p = 0.15$).

**Gender:** Finally, we analysed the users based on their gender and compared their cross-group errors (see *Supplementary Figure 5*). Even though Welch's $t$-test with unequal variances confirms that this difference between males and females in the control condition is indeed significant ($t(349.86) = -2.412$, $p = 0.016$) meaning that female users tend to make predictions that are closer to the overall median answer compared to male users in the control conditions, both the visual analysis and Welch's $t$-test for the same comparison in the treatment condition show that there is no difference among male and female individual biases ($t(194.92) = -0.929$, $p = 0.354$). Thus, both sexes appear to be equally susceptible to the anchoring stimuli provided in the treatment condition.

## Discussion

Over a period of one month, 549 participants made numeric predictions to over 62 questions pertaining to the outcome of a variety of future events after having been exposed to different numeric anchors. Overall, 13,000 predictions were recorded during the experiment. After defining and applying a consistent normalisation method to the distributions of answers of all questions, the various anchoring stimuli and responses were compiled to form the anchoring response function.

The calculation of the anchoring index (AI) proposed by Jacowitz and Kahneman (1995)[1] results in a mean AI of 0.61 for all questions. This means that participants did not fully accept the presented anchoring stimulus in their predictions (which would be the case if AI = 1) but instead adjusted their predictions to a level of roughly 60 percent of the size of the anchor. This effect is in accordance with the one observed by Jacowitz and Kahneman (1995), who found a mean AI of 0.49 (with slightly different effects for low and high anchors). This large size of the anchoring index here is likely attributable to the specific nature of the provided anchors: Firstly, the numeric hints usually contained information about the size of the target outcome in the past, making the anchors appear somewhat relevant to the question. Secondly, the anchor value and target outcome were typically provided in the same unit, such as °C, USD, etc. This provides support for Mochon and Frederick's (2013) scale distortion theory, which proposes that anchors are effective when the stimulus and response share the same scale. Here, further research should be conducted by reducing the relevance of the information provided in the hints even further and varying the units.

The relationship between the anchoring stimuli and responses we find advances the result found by Chapman and Johnson (1994), who suggested that a limit of the anchoring effect exists when stimuli become too extreme. We see a monotonic, yet deaccelerating increase in response curve, and when we modify the measured response by considering the standard deviation of the predictions, we see a decline in the modified response curve. This result is confirmed in the present work by the modified anchoring response function, which resembles the shape of an inverted U. By incorporating the standard deviation of predictions in the treatment groups as a normalisation factor it was observed that larger anchoring stimuli not only translate to larger responses but also lead to higher collective uncertainty among participants.

This finding suggests that the limit of the anchoring bias finds expression in a higher variation of participants' answers rather than in a smaller response. The greater diversity of predictions may be explained by both the original anchoring-and-adjustment, and the selective accessibility theory: Participants may have started an insufficient adjustment process from the implausible anchor, or searched for information confirming the value presented in the anchor. The lack of clear evidence for the presence of one of these mechanisms may suggest that both theories are applicable in this case. Further lab experiments could shed more light on this.

[1] Jacowitz and Kahneman (1995) compute the AI as follows: $AI = \frac{Median\ (high\ anchor) - Median\ (low\ anchor)}{High\ anchor - Low\ anchor}$.

On the level of the individual user it was shown that all participants in the treatment groups were equally affected by the anchoring stimuli in their responses. This holds true regardless of participants' level of engagement during the experiment, their previous performance on the app, or their gender.

One of the implications of this study can be in marketing, where the anchoring effect has already been discussed in the context of pricing (Northcraft & Neale, 1987). Following our results, an optimal pricing strategy might be to first produce a distribution based on prices suggested by a number of different individuals. After cleaning the distribution and removing the outliers, calculate the median and standard deviation of the population, and then opt for a price that is 2-5 standard deviations higher than the median of the distribution. This proposition however needs proper testing through further experimentation.

We must note that our work has a limitation that is mostly due to the experimental design: the anchors provided to the participants in the form of hints had to include relevant information in order to keep users engaged on the app, whereas, Chapman and Johnson (2002) argue that proof for the presence of the anchoring bias is strongest when the information influencing participants' answers is uninformative because subjects had no rational reason to follow that value. The present experiment uses two consecutive anchors that both contain relevant information, however, users have to "pay" to unlock the second hint. Hence, participants' predictions might be biased towards the treatment hint partly because they implicitly (and irrationally) associate a higher importance with it (endowment effect). In order to counteract this effect as much as possible, the information provided in the second hint in many questions was even less pertinent for answering the searched for target outcome than the first hint, or equally uninformative. Thus, the anchoring bias observed in this experiment is due to *overly influential* factual information (Chapman & Johnson, 1994).

A major contribution of this experiment lies in the ecological validity of this research. Not only does the overall setting of the experiment in the form of a mobile application closely resemble a real-world situation in which information is provided by various sources and subsequently processed by the individual, but the various different questions also reflect realistic forecasting decisions.

The fact that the anchoring effect was especially strong for questions in which the anchoring stimuli were labelled as 'PTF-prediction' values constitutes a very important finding of this experiment. This suggests that people's perception of the usefulness or veracity of information can be influenced by the provision of an "authority" label. In an era in which clickbait and fake news are frequently used tools to compete for people's attention and influence their opinions, it is crucial for individuals to understand the strong impact of purposefully (or even unintentionally) set anchors on their perception of certain events: the slogan "We send the EU £350 million a week – let's fund our NHS instead" that was used by the *Leave*-campaign in Brexit vote in 2016, is a prime example. Further research should investigate whether anchors with the same numerical value that are associated with different sources produce distinct anchoring effects. Moreover, given the specific experimental platform used in this work, it would also be interesting to determine how information about predictions made by other users on the app influence participants' predictions as part of a 'collective intelligence' experiment in a future work.

# Methods

## Experimental design

The platform we used for the online field experiment is an application (app) for smartphones called "Play the Future" (PTF), which has been in operation since the beginning of 2016. It is freely available for both Android and iOS devices.[2] Over 300,000 users are currently registered on the app, with around 82,000 monthly active users making numerical predictions regarding the outcomes of various economic, social, sports, entertainment, and other events. For each question two hints are provided that contain relevant information. The first hint is always shown, whereas the second hint can be unlocked with a 'key'. Keys automatically renew themselves for free every four hours so that up to a maximum of three keys are available at any given time.

Participants were not specifically recruited for this experiment; all users who were active on the PTF application during the time period of the experiment (May-June 2018) were treated as participants. The experimental questions were both thematically and structurally very similar to regular PTF questions (apart from the distinction between the two user groups), thus, users were not explicitly notified about taking part in an experiment, which adds to the real-world character of the research. Users were not paid for their participation (the game is free to use); they used the app out of their own volition and could discontinue playing at any point.

The experimental treatment is administered in the second (locked) hint: Its description and content is different for two user groups A and B. By recording each unlocking of the second hint it can precisely be determined which participant was exposed to the treatment and who belongs to the control group on a per-question basis. Participants who have only seen the first hint are considered to be in the control group. The information provided in the first hint is identical across all users. The second hint reveals either a low or a high value (depending on the group) of the target outcome during a specific time frame. We devised the questions and hints in a way that no experimental group has an advantage over the other group.

The two second hints contain values that are above and below the value shown in the first hint. Ideally, the distance to the medium value is roughly the same for both groups, however, due to the hints containing real-world factual information this was not possible in all cases. While generally, the hints are factually accurate, a number of questions include a fictitious "PTF's prediction" value in the second hint to explore whether users are equally anchored by facts and predictions from unverified sources. Additionally, some questions contain irrelevant information in the hint for the control group in order to test whether that information also serves as an anchor. The current events and factual information for all questions and hints were manually researched by the authors. Please refer to *Supplementary Table 1* for a list of all questions.

Out of the six to ten questions that were asked on the app every day, two to four questions were part of the experiment and the remaining questions were regular, non-A/B testing questions designed by the PTF team. Each question was available for a limited time period of approximately 24 to 48 hours. Within that period users could freely decide whether and when to submit a response, thus, time pressure was minimal. We did not control for whether users gathered additional information (e.g., consulting weather statistics) before making a prediction. Users received an accuracy score and points proportional to that two to four days after the results were in.

At the beginning of the experiment, all the registered users were randomly allocated to two groups A and B and new users were subsequently assigned to one of the two groups. The second hints were designed in a way that each participant would receive a roughly equal number of high and low second hints in alternating order. Participants decided on a per-question basis whether to unlock the information contained in the second hint or not given their limited number of keys. This self-selection into the treatment or non-treatment condition is characteristic for online field experiments (Parigi, Santana, & Cook, 2017). As users need to log in to play the game, we could keep track of their actions and the various treatments, as well as their gender and country.

[2] The app is currently unavailable in Europe, due to adjustments being made to accommodate the General Data Protection Regulations which took effect on May 25, 2018.

### Ethical approval

Ethical approval for this experiment was obtained from the Social Sciences and Humanities Interdivisional Research Ethics Committee (IDREC) of the University of Oxford (CUREC 2 reference number: R52154/RE001). The users had agreed to the terms and conditions laid down by PTF upon signing up to the app. Moreover, since the questions in the experiment were in line with previous questions asked on the app, participants could not notice the difference between the experimental and normal setting. At no time were participants exposed to any risks; no deception took place and the questions only pertained to events for which the results will be publicly known and that were appropriate for all ages. As a further precaution, the data collection was kept to an absolute minimum, and any personal information was anonymised by the PTF team before it was released to the researchers for analysis, meaning that no user was individually identifiable.

### Participants

Overall, 549 users took part in the experiment by answering at least one of our questions. One third of participants only answered five questions or less. Roughly one quarter of users answered 80 percent of all experimental questions or more. 56% of users identified as female, 41% as male, and 3% did not disclose their gender. Users are predominantly from North America (60% from Canada and 36% from the US).

## Data preparation and analysis

### Outlier removal and key statistics for each question

The visual inspection of the three answer distributions (control group, group A, group B) showed that all groups contained outlier values. Extreme values can bias a distribution's key statistics so that the calculated and standard deviation are not representative of the data's actual distribution, thus necessitating their exclusion. Data points were classified as outliers and consequently removed if they lay outside the range of median ± 2.5 times the median absolute deviation (MAD).

For subsequent analyses (see below), the median prediction value was calculated for each of the three groups based on the cleaned dataset. Even after removing the most extreme values from the dataset the distributions may still be skewed. Thus, the median rather than the mean prediction serves as a better basis when comparing the various groups because it is more robust to skewness in distributions and therefore representative of participant's collective prediction value in each group (Jacowitz & Kahneman, 1995).

### Curve fitting

The anchoring response curve is the result of plotting a smooth curve through the resulting data points. A non-parametric model relying on local least squares regression (LOESS) was used to create the curve, thus eliminating the need to determine the specific form of the model before fitting (Cleveland & Devlin, 1988; Fox & Weisberg, 2011). This method is particularly suitable for data exploration. The shaded area around the LOESS curve indicates the 95 percent confidence interval.

### Error Index

Users' previous performance on the app is indicated by an error index. The error index is based on the accuracy of the last 10 predictions (or fewer depending on how many times a given user had previously played on the app) before the experiment started. For each of those questions the absolute difference between the user's prediction and the actual result was calculated and normalised by the standard deviation of all predictions for that question according to *equation (7)* (after having removed extreme answers as described above). Subsequently, the median of all calculated error scores per participant was computed. A lower error score is indicative of a better previous performance.

$$(7) \qquad Error\ score\ per\ participant\ i\ per\ question = \frac{|\ Prediction_i - Actual\ result\ |}{\sigma_{All\ predictions\ (cleaned)}}$$

## Acknowledgment

We thank the Canadian company Play the Future, Inc.[3] that has kindly provided the app for research purposes. We also thank Myrto Pantazi for comments on the experimental design, David Sutcliffe for valuable comments on the manuscript, and Mahdi Nasiri for help with analysis. TY was partially supported by The Alan Turing Institute under the EPSRC grant EP/N510129/1. The funder had no role in the conceptualization, design, data collection, analysis, decision to publish, or preparation of the manuscript.

## Author Contributions

TY outlined the study and designed the experiment. JR generated the experiment questions and completed data collection and analysis under the supervision of TY. TY and JR wrote the manuscript.

## Competing Interests statement

The authors declare no competing interests.

[3] Link to their website: http://www.playthefuture.com

# Supplementary Information

## Fooling with facts:

## Quantifying anchoring bias through a large-scale online experiment

Taha Yasseri[1,2]* and Jannie Reher[1]

[1]Oxford Internet Institute, University of Oxford, 1 St Giles, Oxford OX1 3JS, UK

[2]Alan Turing Institute, 96 Euston Rd, London NW1 2DB, UK

*Corresponding Author: taha.yasseri@oii.ox.ac.uk

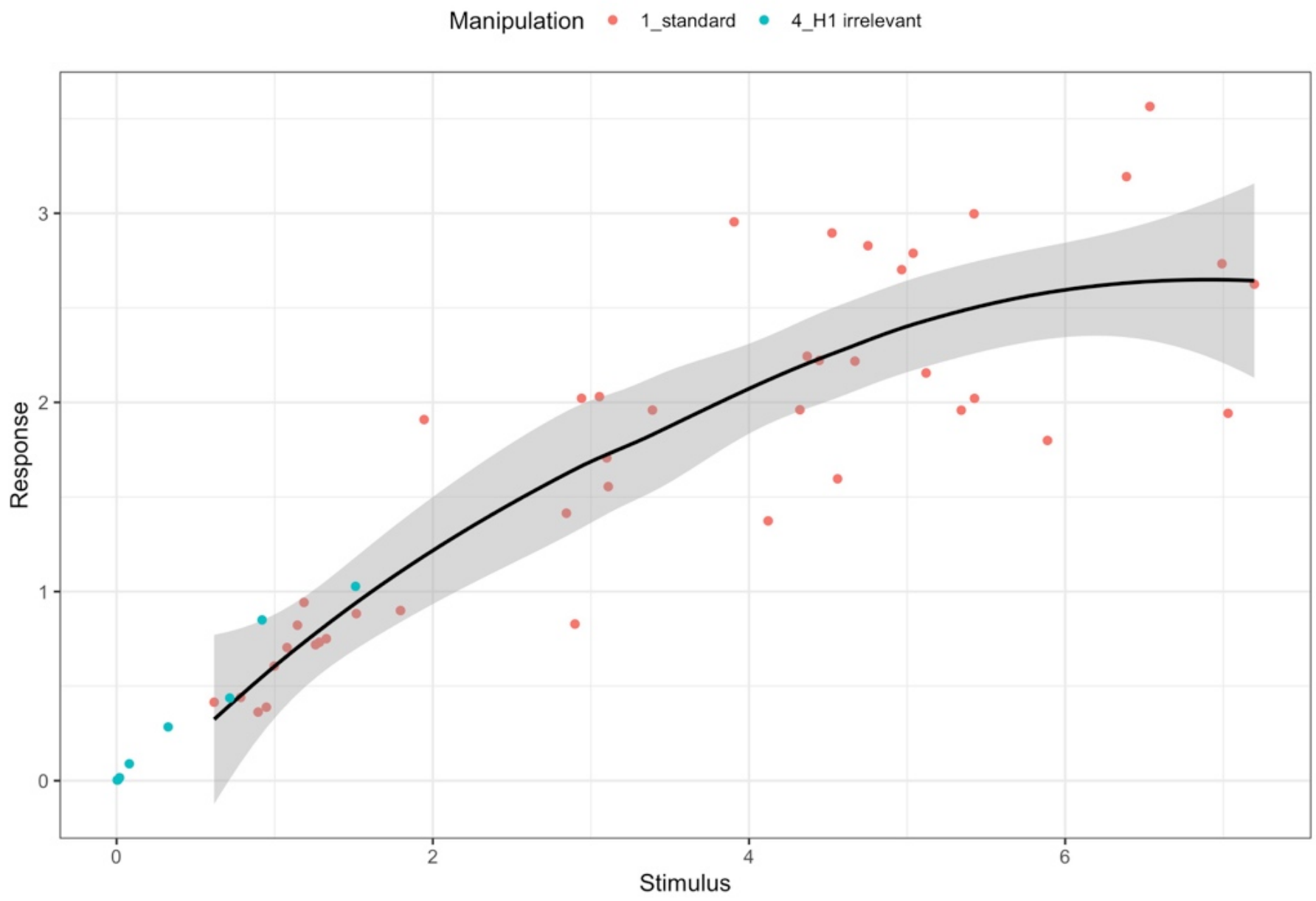

**Supplementary Figure 1.** Response vs. stimulus based on 42 standard questions and 8 questions with an irrelevant anchor in the control group. Points coloured according to manipulation. LOESS curve fitted to standard anchoring questions for reference.

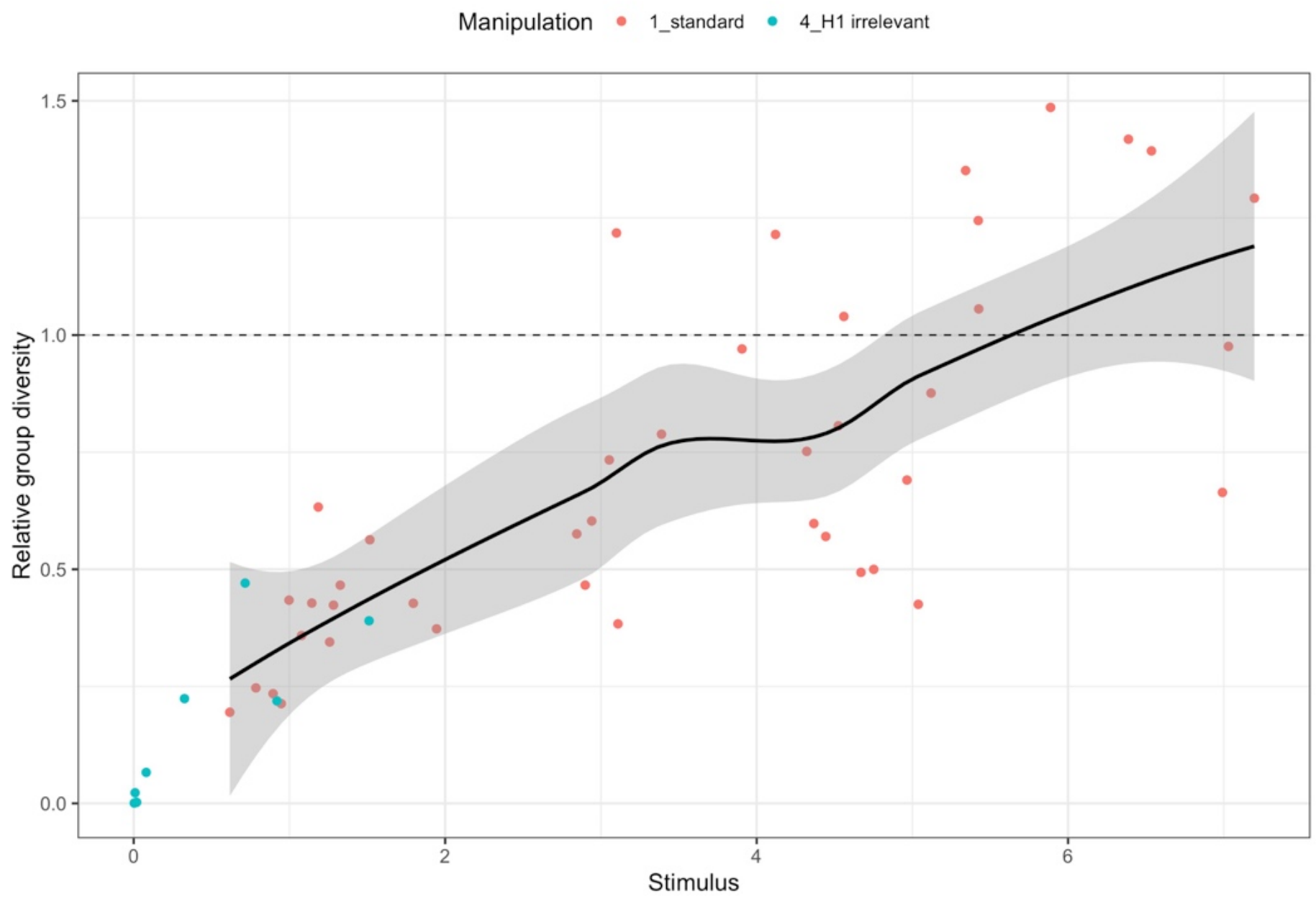

**Supplementary Figure 2.** Relative group diversity vs. stimulus based on 42 standard questions and 8 questions with an irrelevant anchor in the control group. Points coloured according to manipulation. LOESS curve fitted to standard anchoring questions. Line at y = 1 indicates the equal ratio of the treatment to the control group's standard deviation.

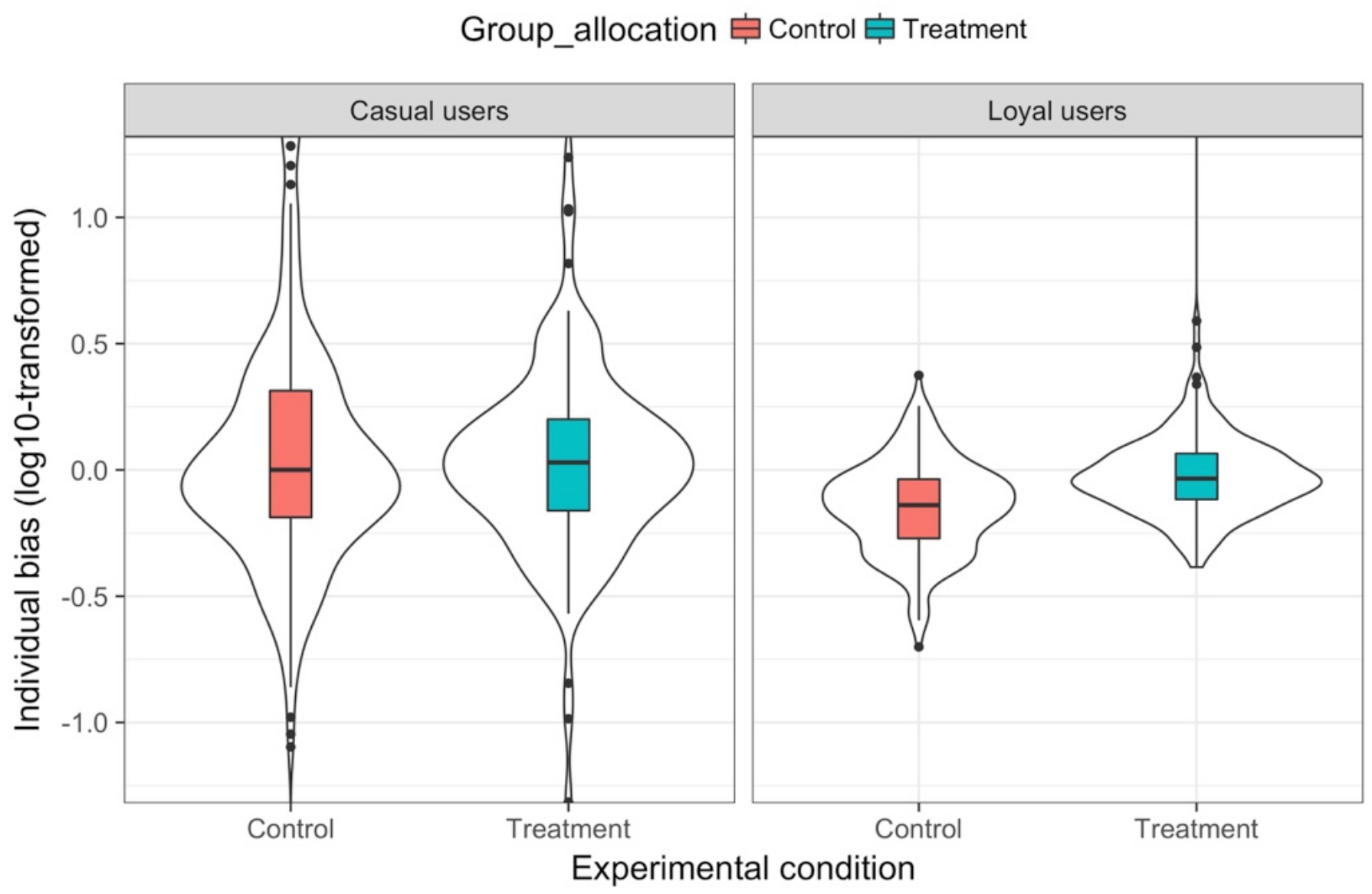

**Supplementary Figure 3.** Individual bias (log10-transformed) vs. experimental condition, distinguished by level of activity during the experiment.

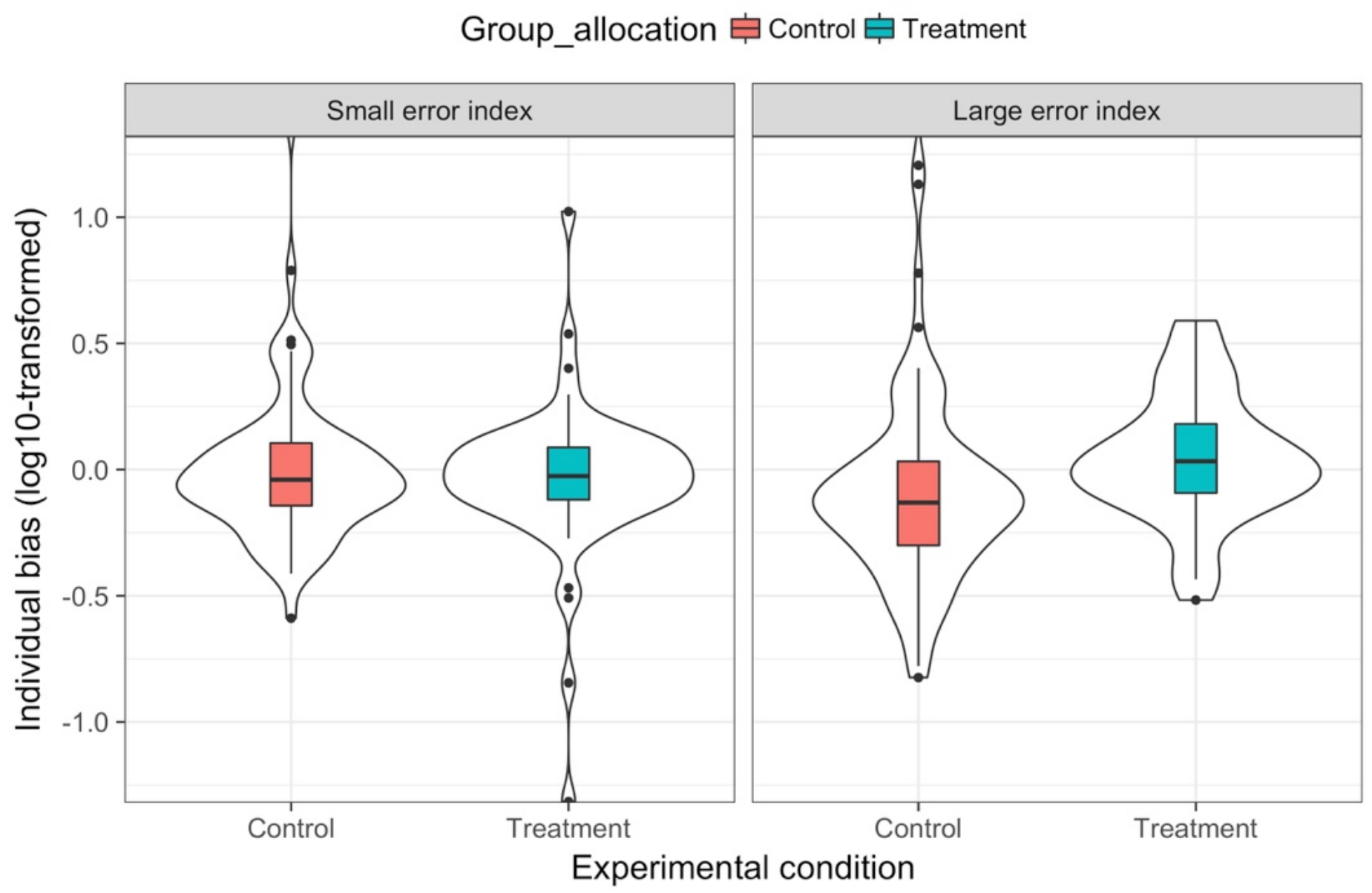

**Supplementary Figure 4.** Individual bias (log10-transformed) vs. experimental condition, distinguished by previous performance on the app.

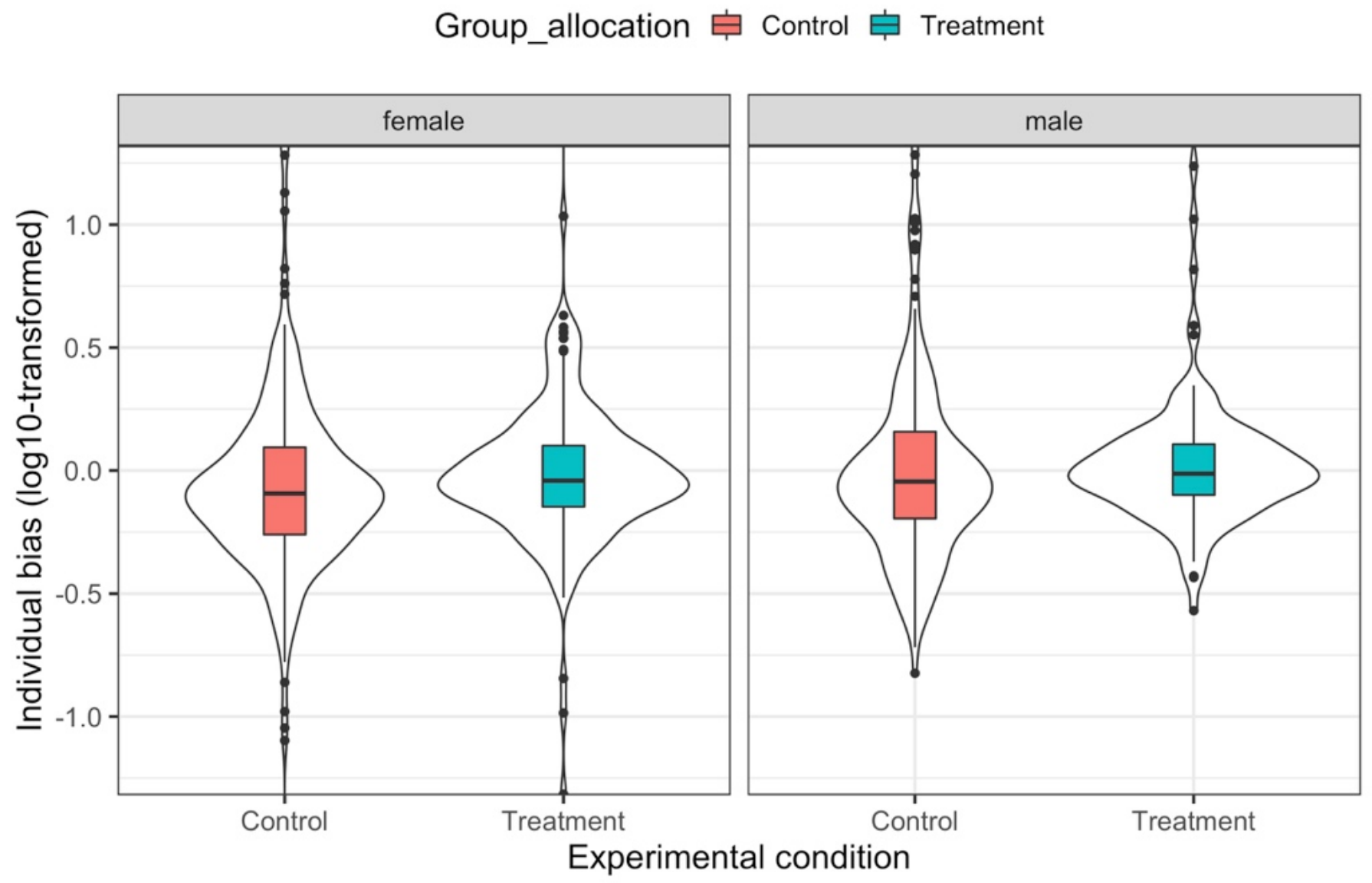

**Supplementary Figure 5.** Individual bias (log10-transformed) vs. experimental condition, separated by gender.

**Supplementary Table 1. Overview of all experimental questions.**

| | | | | Control Group | Experimental Group | | |
|---|---|---|---|---|---|---|---|
| **#** | **Type of question** | **Event name** | **Question** | **Hint 1** | **Hint 2 - Group A** | **Hint 2 - Group B** | **Result** |
| 1 | Standard | Apple WWDC | Apple's Worldwide Developers Conference will take place next week in San Jose where exciting new software and devices may be presented! What will the starting price be in US $ of the device that is announced last during Apple’s WWDC 2018? | Current starting price of Apple’s iPhone 8: 699 USD | Current starting price of Apple’s Watch Series 3 (Cellular): 399 USD | Current starting price of Apple’s MacBook Air: 999 USD | 0 USD |
| 2 | Standard | Apple WWDC | Investors will look closely at Apple's WWDC to see what's in store for consumers during the next months. What will Apple’s closing stock price be on the first day of the WWDC on June 4th? | Average opening stock price during past 3 days: 187.51 USD | Highest stock price yesterday: 188.23 USD | Lowest stock price yesterday: 186.14 USD | 191.83 USD |
| 3 | Standard | Apple WWDC | Aaaaand cut! In HH:MM, how long will this year's WWDC keynote address be? | Scheduled length of WWDC 2018 keynote: 2:00 | Length of WWDC 2017 keynote: 2:19 | Length of WWDC 2016 keynote: 2:03 | 2:15 h |
| 4 | Standard | Apple WWDC | What will the temperature be in San Jose at the start of the keynote (Monday, 10 am PDT) in °F? | Temperature last Monday at 10 am in San Jose: 73 | Forecast temperature high on Monday: 79 | Forecast temperature low on Monday: 57 | 66°F |
| 5 | PTF-prediction | Apple WWDC | During the last couple of years, Tim Cook welcomed several speakers of the company to the stage to introduce new software features or products. How many different people will present during the keynote this year? | Number of people on stage during WWDC 2016: 10 | Number of people on stage during WWDC 2017: 12 | PTF's prediction: 8 | 9 people |
| 6 | PTF-prediction | Little Big Town concert | At the beginning of June the band Little Big Town opened the 2018 CMT Music Awards with the first live performance of their new single "Summer Fever". Now that it's summer, do you also have "that flip-flop attitude" they're singing about? How many views will the official video of "Summer Fever" on the VEVO YouTube channel have by next Wednesday (June 18)? | Number of views this Wednesday: 234,700 | PTF's prediction: 534,700 | PTF's prediction: 384,700 | 346,060 views |
| 7 | PTF-prediction | Little Big Town concert | Little Big Town will be playing at the Country Summer Music Festival this weekend in Santa Rosa, CA. Hopefully it'll be sunny - but we're sure the band manages to fire the crowd up either way with their new single "Summer Fever"! In F°, what will the temperature be in Santa Rosa at 1 PM on Sunday? | Temperature last Sunday at 1 PM: 78°F | PTF's prediction: 82 °F | PTF's prediction: 74 °F | 67 °F |
| 8 | PTF-prediction | Ottawa Craft and Beer Festival | Over 250 craft, import, and domestic beers will be presented at Ottawa's Beer Fest this weekend. The entry fee covers 10 drink tickets and a sampling cup. What will the weather be like at 1 PM in Ottawa, Ontario, on Sunday? | Temperature last Sunday at 1 PM: 68°F | PTF's prediction: 76°F | PTF's prediction: 60°F | 84 °F |
| 9 | PTF-prediction | Sugarland - new music video | Although the American country singers "Sugarland" released their song "Babe" back in April, the corresponding music video aired just a couple of days ago. Taylor Swift went back to her old country-roots when co-writing this song. How many views will the video get on Sugerland's VEVO channel on YouTube by next Thursday? | Number of views last Thursday: 4,206,205 | PTF's prediction: 5,006,710 | PTF's prediction: 5,807,215 | 7,219,035 views |
| 10 | PTF-prediction | U.S. Open Golf | The U.S. Open Golf Championship is partnering with some well-known companies during the tournament: American Express, Deloitte, Lexus, and Rolex. What will the closing stock price be in USD of American Express (AXP) at the NYSE on Tuesday? | Closing stock price AXP last Tuesday: 100.73 USD | PTF's prediction: 99.23 USD | PTF's prediction: 102.23 USD | 97.14 USD |
| 11 | PTF-prediction | Sotheby's auction I | The weather in London can be incredibly unpredictable! Luckily, the Sotheby's Impressionist and Modern Art Evening Sale takes place inside, so no need to worry about the weather. In °C, what will the temperature be like on Tuesday at 1 PM in London? | Temperature at 1 PM last Tuesday: 16°C | PTF's Prediction: 20°C | PTF's prediction: 24°C | 23 °C |
| 12 | PTF-prediction | Sotheby's auction I | Sotheby's is on Instagram and Snapchat! For more information on their latest exhibitions, have a look at Sotheby's accounts! How many Instagram likes will their last post on Monday (June 18) get by Tuesday evening? | Average # likes for pics posted on Mondays in June: 3,078 | PTF's prediction: 2,228 | PTF's prediction: 3,928 | 3,110 likes |
| 13 | PTF-prediction | Sotheby's auction I | The auction house Sotheby's was founded in 1744 and has its headquarters in New York, NY. It has salesrooms all over the world: NY, London, Paris, Hong Kong. Did you know it's the oldest company listed on the New York Stock Exchange? In USD, what will Sotheby's (BID) closing stock price be at the NYSE on Tuesday (June 26)? | Closing stock price BID last Tuesday: 59.06 USD | PTF's prediction: 60.31 USD | PTF's prediction: 57.81 USD | 55.84 USD |
| 14 | PTF-prediction | World Cup Other | There are a number of sponsors supporting the holding of this year's FIFA World Cup - McDonald's is one of them. Have you noticed McDonald's FIFA World Cup Fantasy game online where you create your own team? What will McDonald's (MCD) closing stock price be at the NYSE on Friday? | Closing stock price last Wednesday: 166.58 USD | PTF's prediction: 163.48 USD | PTF's prediction: 169.68 USD | 164.55 USD |
| 15 | PTF-prediction | World Cup Other | Anyone travelling to Russia to see a game of the World Cup live? Apart from the FIFA World Cup, Russia has many interesting cities that are worth a visit: Moscow, Saint Petersburg, .... What will $1 USD be worth in Russian Rubles tomorrow at noon ET? | Last Wednesday's exchange rate on OFX: 62.3800 RUB | PTF's prediction: 62.6615 RUB | PTF's prediction: 62.0985 RUB | 63.4662 RUB |
| 16 | PTF-prediction | Royal Ascot | Royal Ascot - the famous British Horseracing Event - will take place from June 19 to 23. Did you know 1,200 kg of Cornish Clotted Cream will be consumed by the spectators during that time? How many times will the letter "a" be used in Royal | # letter "a" used by @Ascot in last tweet last Saturday: 6 | PTF's prediction: 10 times | PTF's prediction: 14 times | 5 times |
| | | | Ascot's last tweet (@Ascot) on the final race day (Saturday, 23rd)? | | | | |
| 17 | Standard | On this day | A well-known British royal was born on June 21 in 1982: Prince William, Duke of Cambridge will be turning 36 today! We hope he'll enjoy the day with his wife, Catherine, and the kids George, Charlotte, and Louis.<br>How many words will the last tweet tweeted (incl. retweets) by @RoyalFamily by end of day Thursday contain? | # words in @RoyalFamily last tweet last Thursday: 28 | # words in shortest tweet during past 2 days: 10 | # words in longest tweet during past 2 days: 46 | 31 words |
| 18 | Standard | Taylor Swift in London | We hope it won't rain on the day of the first Taylor Swift concert in London - but judging by the "rain show alert" post on Instagram after her Chicago show this "special effect" made the show even more magical!<br>In °C, what will the temperature be at 9 AM on London on Friday (June 22)? | Temperature last Friday at 9 AM: 16°C | Forecast low for Friday: 11°C | Forecast high for Friday: 20°C | 15 °C |
| 19 | Standard | Taylor Swift in London | According to the set list from her first concert in Glendale, Arizona, the show will start with Taylor Swift's hit "... Ready For It". Even of you can't make it to one of her concerts, you can watch the music video of that song and many more on YouTube.<br>How many daily views will Taylor Swift's YouTube VEVO channel have on Monday (June 25)? | Daily average # views past 30 days: 4,951,760 | Smallest # of daily views during last 14 days: 4,623,075 | Largest # of daily views during last 14 days: 5,583,663 | 4,723,825 views |
| 20 | Standard | Big Five Marathon | Pretoria is one of the larger cities very close to the Entabeni Game Reserve where the Big Five marathon will take place. Did you know that Pretoria is South Africa's executive capital?<br>In °F, what will the temperature be like at 11 AM in Pretoria on Saturday? | Temperature at 11 AM last Saturday: 60°F | Forecast low for Saturday: 50°F | Forecast high for Saturday: 68°F | 67 °F |
| 21 | Standard | Big Five Marathon | Have you ever imagined running next to rhinos, zebras, antelopes, and giraffes? Well, you can: at the Big Five Marathon that is taking place on Saturday in South Africa! We bet that's going to be a very adventurous sport event!<br>How many runners will finish this year's Big 5 marathon? | Average number of finishers during last 5 years: 126 | Highest # of finishers during the last 5 years: 152 | Lowest # of finishers during the last 5 years: 101 | 127 racers |
| 22 | Standard | Big Five Marathon | Need some extra motivation to run faster? Well, if a lion chases you, you'd better hurry up! In HH:MM, what will the finishing time be of the runner coming in 80th place at the Big Five Marathon? | Average time of finishers 2017: 05:46 h | Finishing time of runner in 60th place 2017: 05:34 h | Finishing time of runner in 100th place 2017: 06:13 h | 06:14 h |
| 23 | Standard | FORMULA 1 Pirelli Grand Prix De France 2018 | The Formula 1 Pirelli Grand Prix de France will take place at the Circuit Paul Ricard on Sunday, June 24. The French Grand Prix is one of the oldest motor races in the world! It's already been 10 years since the last French Grand Prix took place, so we're excited for the results!<br>In °C, what will the weather be like in Marseille, France, at 11 AM on Monday? | Temperature at 11 AM last Monday: 25°C | Temperature high last Monday: 30°C | Temperature low last Monday: 20°C | 24°C |
| 24 | Standard | FORMULA 1 Pirelli Grand Prix De France 2018 | In Formula 1, a lot of the races are frequently sponsored by large companies. For the upcoming GP de France, the Italian-based company Pirelli is the sponsor. Pirelli is a producer and supplier of tires.<br>In EUR to 2 decimal places, what will Pirelli's (PIRC.MI) stock price be at 1 PM on Tuesday on the Milan Stock Exchange? | Stock price at 2 PM last Tuesday: 7.33 EUR | Stock price low last Tuesday: 7.19 EUR | Stock price high last Tuesday: 7.33 EUR | 7.25 EUR |
| 25 | Standard | City of Vancouver | The City of Vancouver has an official Twitter account. If you're a Vancouver resident, it might be worthwhile checking out the account regularly as useful information and the latest news is announced here on a daily basis.<br>How many words will the last tweet tweeted by end of day Wednesday by @CityofVancouver contain? | # words in @CityofVancouver tweet last Wednesday: 47 | # words in shortest tweet during past 2 days: 3 | # words in longest tweet during past 2 days: 45 | 29 words |
| 26 | Standard | IMVDb | On IMVDb - the Internet Music Video Database - you can discover the latest music video releases of your favorite bands or rewatch the best music videos from the past years.<br>In MM:SS, what will be the length of the newest YouTube video that is published by Thursday evening (if multiple, average will be taken)? | Average length of new music videos last Thursday: 3:57 mins | Length of longest video in past 5 days: 4:30 mins | Length of shortest video in past 5 days: 2:31 mins | 3:37 mins |
| 27 | Standard | Celine Dion World Tour | Céline Dion's 2018 World Tour will kick off on Tuesday at the Tokyo Dome in Japan! This will be her first concert in Asia since her tour in 2008. How many songs will Céline Dion play during the first concert of her Céline Dion Live 2018 tour in Tokyo on June 26? | Average # songs played during 2017 tour: 23 | # songs in longest concert in 2017 tour: 28 | # songs in shortest concert in 2017 tour: 21 | 19 songs |
| 28 | Standard | Auction: Sotheby's Contemporary Art Day Auction | Interested in buying an artwork at one of Sotheby's auctions? It's likely you'll have to dig deep into your pockets - for some pieces the estimated prices are only available upon request! To the 4th decimal place, how many USD can you buy with 1 GBP on Thursday at 1 PM London time? | Last Thursday's exchange rate at 1 PM London time: 1.3215 | Exchange rate low last Thursday: 1.3101 | Exchange rate high during last Thursday: 1.3271 | 1.3094 USD |
| 29 | Standard | Ed Sheeran concert | Did you notice Ed Sheeran album's have very interesting short names? His latest album has the name ÷ (divide) and previous ones are called + (plus) and x.<br>What will be the absolute difference in subscribers on EdSheeran's official YouTube channel on Sunday compared to Saturday? | Daily change last Sunday (absolute): 24,844 | Lowest daily absolute change past 14 days: 20,929 | Highest daily absolute change past 14 days: 29,287 | 28,210 subscribers |
| 30 | Standard | Ed Sheeran concert | This week, Ed Sheeran plays three concerts in The Netherlands and Belgium as part of his current world tour. He'll be coming to North America in August and there are still tickets available!<br>In °F, what will the temperature be in Amsterdam, Netherlands, at noon on Saturday, June 30? | Temperature last Saturday at noon in Amsterdam: 61°F | Forecast high for Saturday: 81°F | Forecast low for Saturday: 56°F | 73 °F |
| 31 | Standard | Elon Musk Birthday | Elon Musk is celebrating his birthday this week: he was born on June 28 in 1971. He is a forward-thinking businessman known for his companies SpaceX, Tesla, Inc. and his latest project, Hyperloop, a high-speed transportation system.<br>In USD, what will Tesla's (TSLA) stock price be at the Nasdaq Stock Market next Tuesday (July 3) at 1 PM? | TSLA's stock price last Tuesday at 1 PM: 334.34 USD | TSLA's stock price low last Tuesday: 326.00 USD | TSLA's stock price high last Tuesday: 343.55 USD | 310.86 USD |
| 32 | Standard | Box Office Mojo | Have you been to the movies lately? Jurassic World 2, Incredibles 2, Ocean's 8, and Solo: A Star Wars Story are some of the films that have been popular during the last month. In USD, how much money will the Top 10 films gross next Monday according to Box Office Mojo? | Top 10 gross last Monday: 30,407,642 USD | Highest gross amount Top 10 last 5 Mondays: 42,991,006 USD | Lowest gross amount Top 10 last 5 Mondays: 10,973,333 USD | 24,575,462 USD |
| 33 | Standard | Formula 1 Eyetime Großer Preis von Österreich | In Formula 1, minimising the time spent for pit stops is essential in order for a driver not to lose his position in the race. On the Formula 1 website a pit stop is even compared to a ballet performance... To the closest second, how much time will the the Formula 1 Austrian Grand Prix winner spend on pit stops? | Average pit stop time all racers last GP: 26.34 secs | Total pit stop time of winner in Australian GP: 21.79 secs | Total pit stop time of winner in Bahrain GP: 24.52 secs | 21.21 secs |
| 34 | Standard | Canada Day | It's July 1 which means it's Canada Day! 🇨🇦🎂 Did you know there is a Twitter account specifically dedicated for Canada? Check it out!How many words will the last tweet tweeted by @Canada by end of day Monday contain? | # words in @Canada's last tweet last Monday: 38 | # words in longest tweet during last 10 tweets: 43 | # words in shortest tweet during last 10 tweets: 11 | 11 words |
| 35 | Irrelevant anchor in control group | Montreal en Arts | The Mtl en Arts festival takes place until the end of this week. This is already the 19th time this event is taking place in Montréal! Did you know this is the largest open-air exhibition of art in Eastern Canada? How many times will the last post by Tuesday evening on Mtl en Art's FB page be liked (incl. other reactions) within 24 hours? | Number of people following Mtl en Arts FB page: 8,379 | Lowest number of reactions last 10 posts: 4 | Highest number of reactions last 10 posts: 73 | 37 reactions |
| 36 | Irrelevant anchor in control group | World UFO Day | On July 2 it's World UFO Day! On this day, people all over the world are on the lookout for unidentified flying objects (UFOs). Have you ever spotted something unusual in the sky? How many upvotes will the last post posted by Tuesday evening in the r/UFOs subreddit on Reddit get within 24 hours? | Number of subscribers to r/UFOs: 101,000 subscribers | Highest number of upvotes last 10 posts: 107 | Lowest number of upvotes last 10 posts: 0 | 0 upvotes |
| 37 | Irrelevant anchor in control group | Paris Fashion Week | The Haute Couture Fashion Week takes place in Paris from July 1 to 5. The next big Fashion shows in New York, London, Milan, and Paris will take place in September when the Spring/Summer 2019 collections will be shown.<br>How many times will the letters V, O, U, G, and E appear in the last tweet tweeted by @VogueParis by Tuesday evening? | Number of tweets overall last Tuesday: 42 | Lowest number of V, O, U, G & E letters in last 10 tweets: 32 times | Highest number of V, O, U, G & E letters in last 10 tweets: 7 times | 18 times |
| 38 | Standard | Wimbledon | It's time for Strawberries and Cream: The Championships, Wimbledon, start on July 2nd. During last year's matches, more than 150,000 merchandise items were sold. Among those were a whopping 8,882 umbrellas and 1,334 panama hats.<br>In °C, what will the weather be like at 11 AM on Thursday in Wimbledon? | Temperature last Thursday at 11 AM in Wimbledon: 18°C | Lowest temperature last Thursday: 12°C | Highest temperature last Thursday: 27°C | 24 °C |
| 39 | Standard | Fourth of July | May the fourth be with you! This day in 1776, the United States of America adopted the Declaration of Independence. The famous Macy's 4th of July Fireworks in New York City are annually televised on NBC - or, if you're in NYC, you can watch them live from the East River. In USD to 2 decimal places, what will Macy's (M) stock price be at 1 PM on Thursday on the NYSE? | Macy's stock price last Thursday at 1 PM: 38.14 USD | Macy's stock price high last Thursday: 38.62 USD | Macy's stock price low last Thursday: 37.51 USD | 36.76 USD |
| 40 | Standard | IMVDb | You can check out the most popular music videos that have been released in the last five days on IMVDb.com! Right now, the song Narcos by Migos leads the list. It's followed by Charli XCX's new video for 5 In The Morning.<br>In MM:SS, what will be the length of the newest YouTube video that is published by Sunday evening? | Length of last Sunday's video: 3:41 mins | Length of shortest video in past 5 days: 2:27 mins | Length of longest video in past 5 days: 5:37 mins | 3:48 mins |
| 41 | Standard | World Cup: Brazil vs. Belgium | On the official 2018 FIFA World Cup Russia homepage you can find plenty of interesting facts about each team and each game. Did you know that for each game a Man of the Match is chosen?<br>How many passes between players (for both teams, incl. incomplete) will be recorded during the Brazil vs. Belgium match on Friday? | # passes between players during Croatia vs. Nigeria game: 850 | # passes between players during Iran vs. Morocco game: 627 | # passes between players during Saudi Arabia vs. Egypt game: 1012 | 927 passes |
| 42 | Irrelevant anchor in control group | Ironman 70.3 Muskoka | The Ironman 70.3 Muskoka Canada takes place on Sunday in Huntsville, ON, close to the Great Lakes. Did you know that the 70.3 stands for the 70.3 miles (113 km) that is covered by the athletes via swimming, cycling, and running? This is half the distance of a regular Ironman.<br>In HH:MM, what will be the finishing time of the racer coming in 45th place at the 70.3 Muskoka Ironman? | Time of winner of last 70.3 in Mont-Tremblant, QC, Canada: 03:41 h | Finishing time of racer coming in 30th place at last 70.3 in Norway: 04:52 h | Finishing time of racer coming in 60th place at last 70.3 in Norway: 05:19 h | 04:38 h |
| 43 | Standard | On this day | 24 years ago on July 5, Amazon.com was founded by Jeff Bezos. The company has its headquarters in Seattle, Washington. Did you know the logo includes an arrow which indicates everything "from a to z" is available at Amazon.com?<br>In USD, what will be Amazon.com's (AMZN) stock price at the NasdaqGS next Tuesday at noon? | Stock price last Tuesday at noon: 1,704.54 USD | Amazon's stock price low last Tuesday: 1,692.48 USD | Amazon's stock price low high Tuesday: 1,725.00 USD | 1,747.06 USD |
| 44 | Standard | MTL Challenge Dragon Boat | Beat the heat 🔥! Watch the MTL Challenge Dragon Boat race in the Olympic Basin in Montréal this weekend! In addition to the exciting races you'll also be able to see live performances from artists from China and Montréal at the Chinese Cultural Festival.<br>In MM:SS, what will be the finishing time of the 10th boat in the 2000m Community Mixed Final A race on Sunday? | Average time in mixed 2000m in 2017: 10:41 mins | Finishing time of boat in 5th place in same race 2017: 10:30 mins | Finishing time of boat in 15th place in same race 2017: 11:07 mins | 11:29 mins |
| 45 | Standard | Tour de France | This year's Tour der France will start on Saturday and finish on the 29th of July. After each of the 21 stages, differently colored jerseys are given to certain racers, depending on their performance during the day. The overall leader wears a yellow jersey.<br>In °C, what will the temperature be in Paris on Monday at 10 AM? | Temperature in Paris last Monday at 10 AM: 22°C | Forecast high for next Monday: 28°C | Forecast low for next Monday: 18°C | 21 °C |
| 46 | Irrelevant anchor in control group | Formula 1 Rolex British Grand Prix | The Formula 1 2018 Rolex British Grand Prix takes place this weekend at the Silverstone Circuit. The racers will do 52 laps, each being 5.891 km long. What will be the average speed of the person who finishes 10th in the British Grand Prix on Sunday during his fastest lap? | Number of cars racing in the Grand Prix: 20 | Speed (average) of first finisher of last British GP (fastest lap): 234.03 secs | Speed (average) of last finisher of last British GP (fastest lap): 227.20 secs | 226.89 secs |
| 47 | Standard | Montreux Jazz Festival | The Montreux Jazz Festival currently takes place in the French part of Switzerland. This world-famous festival has featured many star musicians in the past. This year, Elton John will perform!<br>To the 4th decimal place, how many CHF can you buy with 1 CAD on Tuesday at 1 PM? | Last Tuesday's exchange rate at 1PM: 0.7551 CHF | Exchange rate high, last Tuesday: 0.7569 CHF | Exchange rate low, last Tuesday: 0.7528 CHF | 0.7562 CHF |
| 48 | Irrelevant anchor in control group | Flight statistics | Going on vacation? Getting to your dream holiday location can be a hassle sometimes when flights are delayed, cancelled or baggage is lost.<br>How many flights will be delayed in the Asia-Pacific area next Wednesday? | Global total delayed flights last Wednesday: 21,994 | Delayed flights in North America last Wednesday: 4,457 flights | Delayed flights in Europe last Wednesday: 7,381 flights | 7,853 flights |
| 49 | Standard | MoMA photography | Going back in time! Historic photographs that have been on exhibition in the Museum of Modern Art (MoMA) New York are now up for sale at Christie's. In many photos you can discover how New York's skyline developed over the 20th century!<br>In USD, what will be the final sale price of the most expensive photo in the auction? | Mean price prediction for 5 most expensive photos: 13,000 USD | Christie's low price prediction for most expensive photo: 12,000 USD | Christie's high price prediction for most expensive photo: 18,000 USD | 100,000 USD |
| 50 | Standard | Tim McGraw & Faith Hill Soul2Soul Tour | After playing 80 extremely successful concerts in 2017, Tim McGraw & Faith Hill are on the road again to continue their Soul2Soul World Tour. The two singers not only have an incredible chemistry on stage but also off stage! How many daily views will the TimandFaithVEVO channel on YouTube have on Wednesday? | Daily views last Wednesday: 29,418 | Lowest # of daily views during last 14 days: 25,754 | Highest # of daily views during last 14 days: 32,253 | 30,963 views |
| 51 | Standard | Croatia vs. England | The third last match in this year's World Cup (there's still the Third Place Play-Off happening on the 14th) will take place in Moscow in Russia's largest football stadium. Croatia will meet England for the 8th time. How many passes between players (for both teams, incl. incomplete) will be recorded during the Croatia vs. England match on Wednesday? | # passes during Panama vs. Tunisia game: 910 | # passes during Uruguay vs. Saudi Arabia game: 1,111 | # passes during Senegal vs. Colombia game: 712 | 1101 passes |
| 52 | Standard | Flight NYC to Moscow | Fancy seeing the World Cup final live? Moscow is definitely worth a visit with its various sights: It is famous for The Red Square at the heart of the city, the Kremlin, St. Basil's Cathedral, and more. There's a a lot to explore! In HH:MM, how long will it take flight AFL101 to fly from New York JFK to Moscow on Thursday? | Flight duration, last Thursday: 08:14 h | Longest flight duration during 10-day period: 08:30 h | Shortest flight duration during 10-day period: 7:52 h | 08:16 h |
| 53 | Standard | Festival d'été de Québec | Summer time - festival time! The Festival d'été de Québec 2018 takes place from July 5-15. Stars such as The Weeknd, Neil Young, Shawn Mendes, the Foo Fighters, Lorde and many more will perform! The company Bell is one of the major sponsors of this Festival.<br>In CAD to 2 decimal places, what will Bell's (BCE) stock price be at noon on Thursday? | Stock price at noon last Thursday: 53.43 CAD | Stock price high last Thursday: 53.80 CAD | Stock price low last Thursday: 53.22 CAD | 55.87 CAD |
| 54 | Irrelevant anchor in control group | Film release | A couple of exciting films are being released this month: Mission Impossible - Fallout, Mamma Mia: Here We Go Again, and Skyscraper.<br>In Mio. USD, how much will the Top 12 movies gross during the upcoming weekend (July 13-15)? | Gross value highest grossing film last weekend: 76.0 Mio. EUR | Top 12 lowest gross value during last 5 weekends: 112.3 Mio. USD | Top 12 highest gross value during last 5 weekends: 271.3 Mio. USD | 157.3 Mio. USD |
| 55 | Irrelevant anchor in control group | Friday the 13th | The weekend is almost there! This week, Friday is not just any Friday but Friday the 13th. Did you know that the 13th of a month falls at least once on a Friday every year but not more than three times?<br>How many upvotes will the last post on Friday in the subreddit r/IsTodayFridayThe13th receive within 24 hours? | Number of subscribers to the subreddit: 25.200 | Lowest number of upvotes last 10 posts: 510 | Highest number of upvotes last 10 posts: 628 | 77,000 upvotes |
| 56 | Standard | Trump visit UK (July 12-15) | President Trump will embark on his first official visit to the UK from Thursday onwards for four days. He's scheduled to attend a dinner at Winston Churchill's birthplace and meet with the Queen at Windsor Castle.<br>How many words will the last tweet tweeted by @realDonaldTrump by end of day Friday contain? | # words in @realDonaldTrump's last tweet last Friday: 38 | # words in his shortest tweet during last 10 tweets: 9 | # words in his longest tweet during last 10 tweets: 49 | 0 words |
| 57 | Standard | FB App anniversary | The Facebook for iPhone app launched 10 years ago on July 10, 2008. It took Facebook one year to develop the app after previously only offering a mobile version of its website for people's smartphone browser. How times have changed!<br>In USD to 2 decimal places, what will Facebook's (FB) stock price be at 3 PM on Tuesday at the NasdaqGS? | Last Tuesday's stock price at 3 PM: 203.55 USD | Stock price low last Tuesday: 202.26 USD | Stock price high last Tuesday: 204.91 USD | 208.93 USD |
| 58 | Standard | Flight cancellations | Imagine you've got somewhere exciting to go... and your flight gets cancelled! There can be multiple reasons for that: Staff is on strike, the aircraft is not available, or the weather conditions prevent planes from flying. How annoying!<br>How many flights will get cancelled in the Asia-Pacific region next Tuesday? | Flight cancellations in the United States last Tuesday: 326 | Flight cancellations in Europe last Tuesday: 231 | Flight cancellations in North America last Tuesday: 380 | 744 flights |
| 59 | Standard | Twitter birthday | On July 15 in 2006, Twitter was "born". Can you believe people have been tweeting for 12 years now? Did you know that the first tweet from space was sent in 2009?<br>In USD to 2 decimal places, what will Twitter's (TWTR) stock price be at the NYSE at noon on Wednesday? | Stock price at noon last Wednesday:<br>43.74 USD | Stock price low last Wednesday: 42.22 USD | Stock price high last Wednesday: 44.10 USD | 43.12 USD |
| 60 | Standard | World Cup final | The final game in the World Cup takes place this weekend in which France meets Croatia. It will be played in Russia's capital Moscow in the Luzhniki Stadium, which is not only the biggest stadium in Russia but also one of the largest in the whole of Europe!In °C, what will the temperature be like at 9 AM on Thursday in Moscow? | Temperature at 9 AM in Moscow last Thursday: 22°C | Temperature low last Thursday: 16°C | Temperature high last Thursday: 26°C | 21 °C |
| 61 | Standard | Flight time Toronto to Paris | City of love! France is one of the teams playing in the World Cup final match on Sunday. Its capital, Paris, is definitely worth a visit (think Eiffeltower, Notre-Dame, and Louvre!). There's something for everyone: lots of cafés, shopping opportunities, and museums.<br>In HH:MM, how long will it take flight AC880 to fly from Toronto Pearson to Paris CDG on Thursday? | Flight duration, last Thursday: 06:36 h | Longest flight duration during 10-day period: 06:51 h | Shortest flight duration during 10-day period: 06:23 h | 07:04 h |
| 62 | Standard | BBC Proms | The BBC proms - the world's greatest classical music festival - have begun! The word proms stands for "promenade concerts". These concertsy are informal and inexpensive and the audience stands to listen (promming).<br>How many words will the last tweet tweeted by @bbcproms by end of day Thursday contain? | # words in @bbcproms's last tweet last Thursday: 20 | # words in longest tweet during last 10 tweets: 34 | # words in shortest tweet during last 10 tweets: 6 | 14 words |